
\magnification=1200
\baselineskip=36pt
\baselineskip=18pt
\normallineskip=8pt
\overfullrule=0pt
\vsize=23 true cm
\hsize=15 true cm
\voffset=-1.0 true cm
\font\bigfont=cmr10 scaled\magstep1
\font\ninerm=cmr9
\footline={\hss\tenrm\folio\hss}
\pageno=1
\newcount\fignumber
\fignumber=0
\def\fig#1#2{\advance\fignumber by1
 \midinsert \vskip#1truecm \hsize14truecm
 \baselineskip=15pt \noindent
 {\ninerm {\bf Figure \the\fignumber} #2}\endinsert}

\def\ref#1{$^{[#1]}$}
\def\sqr#1#2{{\vcenter{\vbox{\hrule height.#2pt
   \hbox{\vrule width.#2pt height#1pt \kern#1pt
   \vrule width.#2pt}\hrule height.#2pt}}}}

\noindent{preprint HLRZ 23/93}
\bigskip
\bigskip
\centerline{\bigfont LATTICE BOLTZMANN MODELS FOR}
\centerline{\bigfont COMPLEX FLUIDS}
\bigskip
\bigskip
\bigskip
\bigskip
\smallskip
\centerline{{\bf E. G. Flekk\o y$^{1,2}$ and H. J. Herrmann$^{1}$}
{\footnote{$^{\dagger}$}{on leave from C.N.R.S., France}}}
\medskip
\centerline{$^{1}$ HLRZ, KFA J\" ulich, Postfach 1913}
\centerline{5170 J\" ulich, Germany}
\smallskip
\centerline{$^{2}$ Center for Advanced Studies}
\centerline{ Drammensveien 78, Oslo 2, Norway}
\bigskip
\bigskip
\bigskip
\noindent{\bf Abstract} \par
\smallskip

We present various Lattice Boltzmann Models which reproduce
the effects of rough walls, shear thinning and granular flow.
We examine the boundary layers generated by
the roughness of the walls. Shear thinning
produces plug flow with a sharp density
contrast at the boundaries. Density waves are spontaneously generated
when the viscosity has a nonlinear dependence on density which
characterizes granular flow.
\bigskip
\bigskip
\leftline{PACS numbers: 03.40.Gc 47.50.+d 81.35.+k}

\vfill\eject
\noindent{\bf 1. Introduction}
\smallskip
Many fluids in our daily life have rather complex rheological
behavior. Pastes, suspensions, liquid crystals, dense polymers
and granular media are usually non-Newtonian and can exhibit
many flow anomalies, prominent ones being
shear thinning or thickening and the spontaneous formation
of density fluctuations in granular flow. Within the framework
of classical fluid dynamics it is in general
not simple to take into account these nonlinearities.
Therefore it is of
interest to look for alternative techniques to model the
behavior of complex fluids.

As one alternative to the direct solution of the equations of motion
the so called Lattice Boltzmann Models (LBM) have been proposed\ref{1,2}.
These models are defined on a lattice  with velocity vectors that
can only point into few discrete directions and all have the same length.
This simplification is somewhat compensated by the fact that
on each site one has more real degrees of freedom (six on a triangular
lattice) than in the classical numerical techniques allowing for the
definition of a local shear or a local rotation.

Two important questions concerning the LBM models can be asked:
1. How well do they reproduce solutions of the phenomenological
equations of motion, like Navier Stokes? and 2. How well do
they reproduce nature? The first question has been extensively
addressed in the literature\ref{1-6}. If certain assumptions
are made on the length and time scales over which the variables
can change the incompressible Navier Stokes equation can be
derived using the Chapman Enskog scheme. It is, however, known that
straightforward simulations of the LBM can give inhomogeneous
densities violating this incompressibility
restriction. In this paper we will investigate
these spatial and temporal density fluctuations in more detail.
 In fact, we
want to address mainly the second question: Can LB models
describe real phenomena like shear thinning, density waves or
the perturbations arising from the roughness of walls?

For that purpose we will investigate the flow through a pipe
along which the particles are accelerated through gravity.
We want to see what happens if the walls of the pipe are rough
and study constitutive laws that produce plug flow
and clogging.
The typical experimental materials
to which our investigations should apply, are
suspensions with shear thinning in the case of plug flow,
and granular media in the case of clogging.

In the following section we describe the model and the various variants
used in this paper. The next section is devoted to the effects
of wall roughness. Section 4 discusses models that give shear
thinning and section 5 presents data for a model that spontaneously
generates density waves.
\bigskip
\noindent{\bf 2. Description of the model}
\smallskip
We consider a triangular lattice, and on each site $\vec x$ we have six
real variables $N_i(\vec x, t)$, $i = 1,..6$, representing (counted
counter clockwise) the densities of the particles going in the
direction $i$ of the lattice. (For convenience
we will in the following omit the site index $\vec x$ and denote
by $N_i'$ the value of the particle density after collision.)
One updating of the system ( $t \rightarrow t+1$ )
 is given by two steps: (1.) The collision
step at which the six $N_i$ are updated at each site through
$$N_i' = N_i + \lambda (N_i - N_i^{eq})\eqno(1)$$ and (2.) the
propagation step at which each $N_i$ is shifted to the site of
the nearest neighbor in direction $i$.
Eq.~(1) produces a relaxation towards the equilibrium densities
$N_i^{eq}$  which is numerical stable provided the
 relaxation constant $-2 < \lambda < 0$.
The value of $\lambda$ sets the kinematic viscosity of the fluid.
The equilibrium densities are given by $$N_i^{eq} = {\rho \over 6}
(1 + 2\vec u \cdot \vec c_i + 4(\vec u  \cdot \vec c_i)^2
- 2\vec u^2)\eqno(2)$$
where $\rho $ is the mass density at site $\vec x$
$$\rho = \sum_i N_i\ \ ,\eqno(3)$$
$\vec c_i$ the unity vector along direction $i$ and
$\vec u$ the velocity vector at site $\vec x$ defined through
the momentum density per site
$$\rho \vec u = \sum_i \vec c_i N_i\ \ .\eqno(4)$$
The equilibrium distribution $N_i^{eq}$ given in Eq.~(2),
is chosen to give mass
and momentum conservation in the collision step.
The flow will be forced into the direction of the gravity $\vec g$,
which is pointing parallel to the walls of the pipe.
For that purpose an additional step is added after the
collision step which is defined by
$N_i'' = N_i' + {1 \over 3} \vec c_i \cdot (\rho \vec g)$.
Periodic boundary conditions are
imposed in the direction of gravity in which the system has
a length of $L_1$. In the perpendicular direction one has
walls separated by $L_2$ lattice spacings.
The lattice orientation is such that one of the lattice directions is
parallel to the walls.
At the beginning of the simulation the average density
$\bar \rho$ is fixed. It is an important parameter of the
model which because of mass conservation stays constant in time.
We initialize the system by having the same values of
the $N_i$ on each site and then let the system evolve to
its steady state.
In the case of the stable flows steady state is reached after 2000
or 3000 time steps. In the case of the unstable flows that develop
density waves, the simulations might take up to 20000 time steps to
reach steady state.

The sites lying on the walls of the system only have two
directions $a$ and $b$.
Usually two different collision steps can be applied
on these sites\ref{4}, either
the specular condition, i.e. $N_a' = N_b$ and
$N_b' = N_a$, or the bounce-back condition, i.e. $N_{a,b}' = N_{a,b}$.
In the propagation step for these sites the direction in which the
$N_i$ are shifted is inverted. We want to
be able to implement walls that
are not smooth but rugged, i.e. that have (quenched) disorder.
For this purposes we introduce a mixed boundary condition
defined through $$N_a' = x N_a + (1-x) N_b \ \ {\rm and}\ \
N_b' = (1-x) N_a + x N_b\eqno(5)$$ where $x = y^{\alpha}$
and $y$ is a random variable
chosen from a homogeneous distribution between 0 and 1.
Setting $\alpha = 0$ will give the pure bounce-back condition whereas
$\alpha = \infty$ corresponds to the pure specular reflection
condition.

The relaxation parameter $\lambda$ depends on
  the material
properties including the kinematic and the bulk viscosities.
Usually complex fluids are phenomenologically described
by ``constitutive laws'' given e.g. by the functional dependence
of the viscosities upon the shear velocity, the density or the
pressure. We want to investigate the effect
of rather typical nonlinear constitutive laws on the flow
properties. Since an exact relation between $\lambda$ and
the material constants is not known we will lean on some
approximative arguments\ref{1,13} that predict a vanishing bulk
viscosity.
In that case one can relate
$\lambda$ directly to the kinematic viscosity $\nu$ through
$\lambda = -{1 \over 2} (0.25 + 2\nu )^{-1}$. We will consider
two cases: (1.) $\nu$ is a function of the local shear rate $\dot \tau$
and (2.) $\nu$ is a function of the local density $\rho$.
The detailed functional forms used here will be described
in sections 4 and 5.

Our calculations were performed on a Connection Machine CM-2
at GMD (Bonn) using 32-bit precision. The program needs less than one
minute to make $50$ updates of a system of size $1024^2$.
The program was also benchmarked on  a CM-5 at I.P.G. in Paris\ref{7}.
\bigskip
\noindent{\bf 3. The effects of rugged walls}
\smallskip
As already mentioned, it is well known that LB models
produce inhomogeneities in the density  when used
to simulate for instance flow through a pipe.
In the middle of the pipe the density is  higher than
at the walls\ref{13} by
a factor $1/(1-u^2)$ .
This is seen in fig.~1 which shows the density in
a cross section through the pipe. For $\alpha = 0$, i.e. the case
of smooth walls the density profile has precisely the predicted shape
of $\rho_{wall}/(1-u^2)$ as can be seen from the line showing
the pressure \ref{13}   $p = (\rho /2 ) (1-u^2)$.
In the case of rough walls, for which we have chosen in $\alpha = 1$
in fig.~1, the density has a minimum close to the walls.
Also, the pressure
has a minimum at the wall
and a lower, but still constant value in the center. As expected there are
some random fluctuations close to the walls.

In fig.~2 we see the density variation along the center of the
pipe. Since the values taken at different times coincide
very well one is in the steady state.
Clearly the randomness of the reflection properties of the
wall still have some effect but the relative variation is of the
order of 0.0001, i.e. extremely weak.

The roughness at the walls therefore seems to be screened very
efficiently.  This is seen more clearly in fig.~3 which shows the
entire density profile in the pipe.
The boundary  layer has a thickness  of a few mean free paths
(the mean free path in this context is the characteristic length
over which a perturbation in the $N_i$'s will be damped and has
typical length of $1/\lambda$) where  the distribution of the $N_i$'s
is clearly different from that in the bulk of the material.
In this sense it may be characterized as a Knudsen layer.
\bigskip
\noindent{\bf 4. Shear thinning and plug flow}
\smallskip
Shear thinning can be phenomenologically explained by
a non-Newtonian constitutive law given by a
decrease of the viscosity as a function of the shear
rate. Within the context of the LBM the shear rate
$\dot \tau$ can be defined through $$\dot \tau (\vec x)=
{1 \over 3} \vert \sum_i \vec c_i u_{\parallel}
(\vec x + \vec c_i) \vert\eqno(6)$$ where $u_{\parallel}$ is
the projection of $\vec u$ into the
direction of the pipe. We consider a constitutive law
of the form $\nu = \nu_1$ for $\tau \leq \tau_0$ and
$\nu = \nu_2$ for $\tau > \tau_0$ and $\nu_1 > \nu_2$.

In fig.~4 we show the velocity profile in a cross section
through the pipe.
In the simulations the flow was initialized
with a relatively strong forcing, $g = 5 \times 10^{-5}$.
During this initial phase the shearthinned regions
at the walls appear, and the  flow velocity increases
to approximately its steady state value. Then the forcing
was reduced by a factor 10 to  $g = 5 \times 10^{-6}$ and the
system allowed to reach steady state.

 We see that the profile is rather
flat in a broad central region which ends at a sharp
kink after which one finds a rather steep velocity
gradient towards the walls where the fluid is in the thinned
phase.
 This kind of behavior is
usually called plug flow. It was checked that the flow is really
in a steady state by performing longer runs.
In a recent preprint\ref{9} a simulation of a similar
LBM has been presented which also finds plug flow
by taking a constitutive law in which the viscosity
decreases with the shear rate like a power law.
This seems to indicate that the appearance of plug flow
is rather independent on the detailed form of the
constitutive law as long as $\nu$ is a decreasing function
of $\dot \tau$.
\bigskip
\noindent{\bf 5. Density waves}
\smallskip
A salient feature of granular media is the spontaneous
formation of density waves, similar in fact to traffic
jams on highways. One possibility to explain the
effect that generates these waves is to assume that
the viscosity depends on density. Within the
kinetic gas theory of granular media\ref{10,11}
the relation $\nu \propto (\rho - \rho_c)^{1/3}$ has
been derived. Since the above relation imposes a
maximum density $\rho_c$ it is rather difficult
to implement it directly within the context
of the LBM where the particles do not have an
exclusive volume.
 We therefore  chose a piecewise
linear relation of the form $\nu = \nu_{min}$ if $\rho \leq \rho_t$
and $\nu = \nu_0 + \gamma (\rho - \overline{\rho})$
for $\rho > \rho_t$ (see fig.~5).
$\overline{\rho}$ is the average density and the threshold density
$\rho_t$ is chosen to make $\nu$
a positive continuous function
of the density.  Fig.~6 shows results from simulations
where $\rho_t = 2.962$ and the slope $\gamma = 6.25$
corresponding to a minimum
cut-off viscosity $\nu_{min} = 0.01$.

In order to generate density waves we found it necessary
to introduce a small perturbation producing a  0.3\% relative density
difference.
This perturbation was performed by introducing a small
amount of momentum on one line across the pipe,
keeping the mass unchanged. In fig.~6 we see
that this initially very weak perturbation dramatically
builds up and develops into a density wave of over
10\% density contrast.
For a pipe of same width but half the length, i.e. a different
aspect ratio the wave has a less pronounced
profile. This dependence on the aspect ratio is not to
be confounded with finite size effects. Our mean free path
is typically one lattice spacing so that the strong
finite size effects encountered in some lattice gas models\ref{14}
should not be relevant here.
The maximum flow velocity $u_{max}$ at the later times is
$u_{max} = 0.039$ and  $u_{max} = 0.048$ for the
channel lengths 256 and 512 respectively and same width 64.
The forcing $g = 3.33\times 10^{-5}$
is the same for both system lengths.
For the present parameter values the initial perturbation
relaxes, leaving a time independent density field, if
the value of $\gamma$ is less than 3.75.
This effect can be understood qualitatively by
observing that there are two competing mechanisms
in the system: On one hand, the viscous relaxation of density
perturbations will tend to smoothen density contrasts.
On the other hand, the rather steep increase of viscosity with density
combined with the presence of the walls will tend
to increase the contrasts. A small increase in the density
at the wall will give a local increase in viscosity and
 slow down the flow.  Due to the
inertia of the surrounding flow this in turn
will lead to a further increase in the density and so on.
If this instability dominates the relaxing mechanism the
density wave will form.

 By  triggering the
density wave by two spatially separated
perturbations, rather than just a single one,
we checked that the complex shape
of the waves does not reflect the detailed way in which they
were initiated.
We also observe that there seems to be no characteristic
wavelength: Fig.~6 shows that the waves have roughly
the same shape on the scale of the channel length although
the amplitude depends on the system size.

Fig.~7 shows this amplitude as a function of
time during 60,000 time steps.
The insert shows the initial unstable phase
leading to the rather drastic increase of the amplitude
at the time 10,000.
The first small increase in the density
is due to the accelleration of the flow and can be understood
from the velocity dependence in the pressure.
The small jump at
the time 2500 results from  the perturbation.
 Before the instability is triggered at the time 10,000,
small oscillations in the amplitude are observed.
 It was checked that the  amplitude
indeed has its' steady state value at the time 60,000
by running the simulations ten times longer.
The complicated relaxation towards the fully developed
density wave indicates that strong non-linear effects
come into play rendering a linear stability analysis
meaningless. It would be
interesting to understand this behaviour further.

In fig.~8 one can see the density wave propagating.
The fronts are actually sharpest at the boundary and the
left gradient which is less sharp than the right one
has some weak spatial oscillations.
The fact that the waves are of the order of the length
of the pipe again shows that there is no characteristic
length scale. The periodic boundary conditions seem crucial
to reinforce the steady state. One therefore has the typical
behaviour of a kinetic wave as also
found in traffic jam models\ref{15}.
\bigskip
\noindent{\bf 6. Conclusion}
\smallskip
We have presented various versions of Lattice Boltzmann Models
which can reproduce rather complex flow behavior.
On one hand we investigated how the laminar flow
screens the asperities arising from rugged walls by
forming  Knudsen layers close to the wall.
When a shear thinning constitutive law is introduced
we find plug flow.
Finally, when the viscosity is an increasing function
of density we observe a range of parameters for which
the material spontaneously produces density waves
traveling upstream. These density waves are triggered
by some perturbation that apparently is unstable, but
the final shape of the wave is independent of the
initial disturbance.

Plug flow and density waves are common phenomena in
non-Newtonian fluid dynamics and have been investigated
recently in detail for granular media\ref{12}.
It seems therefore that LB models can be a powerful tool
to handle complex fluids numerically.
This approach is, however, yet quite preliminary.
One has to determine the physical
parameters for which the proposed models do match
a real experiment and then compare measured and simulated
results. Work in this direction is in progress.
\bigskip
\leftline{\bf Acknowledgements}
\smallskip
We thank Dan Rothman and Einat Aharonov for illuminating
comments and discussions.
\bigskip
\noindent{\bf References}
\medskip
\item{1.} R. Benzi, S. Succi and M. Vergassola, Phys. Rep.
{\bf 222} (1992), 145
\item{2.} G.R. McNamara and G. Zanetti, Phys. Rev. Lett.
{\bf 61} (1988), 2332
\item{3.} S. Succi, R. Benzi and F. Higuera, Physica D
{\bf 47} (1991), 219
\item{4.} P. Lavall\'ee, J.P. Boon and A. Noullez, Physica D
{\bf 47} (1991), 233
\item{5.} E.G. Flekk\o y, to appear in Phys. Rev. E
\item{6.} A.K. Gunstensen, D.H. Rothman, S. Zaleski and
G. Zanetti, Phys. Rev. A {\bf 43} (1991), 4320
\item{7.} H.J. Herrmann, S. Melin and P. Ossadnik, unpublished report
\item{8.} Y.H. Qian, D. d'Humi\`eres and P. Lallemand, Europhys. Lett.
{\bf 17} (1992), 479
\item{9.} E. Aharonov and D. Rothman, preprint
\item{10.} P.K. Haff, J. Fluid Mech. {\bf 134} (1983), 401
\item{11.} J.T. Jenkins, Arch. Rat'l. Mech. Anal., {\bf 87} (1985), 355
\item{12.} G. Ristow and H.J. Herrmann, preprint; Th. P\"oschel,preprint
\item{13.}  U. Frisch, D. d'Humieres, B. Hasslacher,
P. Lallemand, Y. Pomeau and J. P. Rivet,
 Complex Systems {\bf 1} (1987), 649
\item{14.} G.A. Kohring, Physica A {\bf 186} (1992), 97
\item{15.} K. Nagel and M. Schreckenberg, J. Physique I {\bf 2}
(1992), 2221
\bigskip

\noindent{\bf Figure Captions}
\medskip
\item{\bf Fig.~1} Density as a function of the position $Y$ across the
channel for $\alpha = 0$ (squares) and $\alpha = 1$ ($\times$).
Also $2p = \rho  (1 -  u^2 )$ is shown for
$\alpha
= 0$ ($\diamond$) and $\alpha = 1$ (+). $L_1 = 256, L_2 = 64$,
$\bar \rho = 3$, $g = 3.33\times 10^{-5}$ and $\nu = 0.25$.
The figure shows the steady state profile after 7500 iteration steps.
The maximum flow velocity $u_{max} = 0.52$.
\item{\bf Fig.~2} Density in the center of the pipe
as a function of the  position $X$ along the channel
for $\alpha = 1$ and otherwise the same parameters
as in fig.~1. The two lines correspond to two measurements
25 iteration steps apart.
\item{\bf Fig.~3} Density contrast in the pipe for the same
parameters as in fig.~1 and fig.~2. White denotes the lowest density and
black the largest one.
\item{\bf Fig.~4} Velocity as a function of the  position,
$Y$ across the channel  measured at steady state after
 5000 iteration steps .
The insert shows the viscosity's dependence on the local
shearrate.
In this simulation $\nu_1 = 1.0$, $\nu_2 = 0.1$,
$\tau_0 = 10^{-3}$, $\bar \rho = 3.0$, $ L_1 = 256$ and $L_2 = 64$.
 The forcing is $g = 5\times 10^{-6}$.
\item{\bf Fig.~5} The density dependence of the viscosity chosen in
the simulations.
\item{\bf Fig.~6} The density in the center of the channel as a function
of the  position $X$ along the channel
for $\rho_t = 2.962$,  $\bar \rho = 3.0$,
$g = 3.33 \times 10^{-5}$ and $L_2 = 64$.
The curve of crosses is for $L_1 = 256$ and
60,000 iteration steps after the initial perturbation.
The other curves correspond to $L_1 = 512$ and 5000 (thick line),
60,000 (full line) and 60,025 (dashed line) iterations after
the perturbation was applied. The slope $\gamma = 6.25$ and
the minimum viscosity $\nu_{min} = 0.01$.
\item{\bf Fig.~7} The amplitude, i.e. difference between
largest and smallest density, along the center of the pipe
as a function  of time
measured in units of 100 iteration steps for
$L_1 = 256$ and otherwise the same parameters as in fig.~5.
The insert is a blow-up of the behavior at early times.
\item{\bf Fig.~8} Density contrast in the pipe for the same
parameters as in fig.~5 and fig.~6 after 60,000 (upper) and 60,025 (lower)
iteration steps. White denotes the lowest and black the largest
density.
\end

{}.

{}.